\documentclass[a4paper]{article}
\usepackage[utf8]{inputenc}
\usepackage{listings}
\usepackage{graphicx}
\pdfoutput=1

\newcommand{\keywords}[1]{\textbf{Index terms:}\quad #1}

\title{To Pool or Not To Pool? Revisiting an Old Pattern}
\author{Ioannis T. Christou \\
Athens Information Technology, 15125 Marousi, Greece \\
\and 
Sofoklis Efremidis \\
Athens Information Technology, 15125 Marousi, Greece\\
}

\begin{document}

\maketitle

\begin{abstract}
We revisit the well-known object-pool design pattern in Java. In the last decade, the pattern has attracted a lot of criticism regarding its validity when used for light-weight objects that are only meant to hold memory rather than any other resources (database connections, sockets etc.) and in fact, common opinion holds that is an “anti-pattern” in such cases. Nevertheless, we show through several experiments in different systems that the use of this pattern for extremely short-lived and light-weight memory objects can in fact significantly reduce the response time of high-performance multi-threaded applications, especially in memory-constrained environments. In certain multi-threaded applications where high performance is a requirement and/or memory constraints exist, we recommend therefore that the object pool pattern be given consideration and tested for possible run-time as well as memory footprint improvements.
\end{abstract}


\section{Introduction}
The “Object Pool” design pattern is a well-known design pattern (\cite{wikiobjectpool}) that is often used with so-called “heavy objects” such as database connection objects or thread objects; its usefulness in such cases is immediately justified by the requirement to constrain the number of simultaneous such objects according to the underlying resources available; or by the costs associated with creating and/or destroying such objects. To save such costs, a pool of such objects is created before first use of an object, so that when client code requests an object of the type contained in the pool, the pool is queried for available objects, and if such an object is available it is immediately returned from the pool. When the client code no longer needs the requested object, most usually, the object has to be returned to the pool by invoking an appropriate method; alternatively, the object must have appropriate finalization and/or resurrection methods to guarantee its safe return to the pool. The debate on the usefulness of object pools in managed-memory languages such as Java and C\# (\cite{wikiobjectpool}, \cite{Chen}, \cite{Kircher}) was initially centered around the fact that object allocation and certainly, garbage collection was much more costly for such languages, and therefore object pools were worth the extra effort on the part of the programmer. As better technology was invented for VMs, and in particular, as strong progress was made in the fields of object allocation and garbage collection, object pools lost their appeal, while it remained true that implementing them required significant care on the part of the programmer especially in multi-threaded applications. In the words of J. Bloch~\cite{Bloch}
\begin{center}
“...avoiding object creation by maintaining your own object pool is a bad idea unless the objects in the pool are extremely heavyweight... Generally speaking, however, maintaining your own object pools clutters your code, increases memory footprint, and harms performance. Modern JVM implementations have highly optimized garbage collectors that easily outperform such object pools on lightweight objects.”
\end{center}
Similar advice is offered by B. Goetz~\cite{Goetz}. In the blogo-sphere, almost unanimously, everyone supports this claim (\cite{se}, \cite{dzone}, \cite{so}). Indeed, some attempts at creating fast object pools such as the “furious-objectpool”~\cite{furiousobjectpool} that claim to be faster than standard approaches such as the Apache Commons Object Pool library~\cite{apachecommons} unfortunately suffer from subtle synchronization bugs (e.g., by using the notorious Double Check-Locking idiom), further adding support to the notion that even when implemented correctly, object pools are “synchronization bottlenecks”. However, while optimizing a number of codes for scientific computing applications some of which currently form integral parts of heavy-load production systems, we discovered that under certain conditions, appropriately designed object pools with thread-local object storage for light-weight objects can yield very significant performance benefits. We first discuss an essentially trivial case for using an object pool in the form of a cache data member, then illustrate a more complicated case, and finally introduce our own generic library that we use as solution to the problem. We finally draw some conclusions from this research.

\section{Setting the Stage: Creating Arrays in Random Search}

To set the stage, we first discuss an almost trivial search algorithm that is however often used as a benchmark against which other algorithms’ performance is compared: pure random search! Consider the classical box-constrained non-linear optimization problem $\min_{l\le x \le u}f(x)$. The easiest attempt at solving the problem, is pure random search, also known as Monte-Carlo search, where we evaluate the function f on a given number of randomly selected points in the multi-dimensional box $[l,u]$. A MonteCarloSearch class in an Object-Oriented design, might then implement an \texttt{OptimizerIntf} that would declare a \texttt{double optimize(FunctionIntf f)} method signature (this is the approach taken in the \textit{popt4jlib} open-source library~\cite{popt4jlib}. Since function evaluations are presumably independent of each other, we can easily parallelize the random search by splitting the total number of function evaluations among a number of threads. The class’s implementation of the \texttt{optimize()} method would then create a number of threads that would execute a \texttt{MonteCarloTask} whose \texttt{run()} method includes a loop as shown in Listing~\ref{lst:lst1}.

\begin{lstlisting}[caption={Random vector generator relies too heavily on the garbage collector as it generates a \texttt{double[]} each time the \texttt{newVector()} method is invoked; the array will be used in a single function evaluation and will then be discarded, so it is extremely short-lived.},label={lst:lst1},captionpos=b,frame=single,basicstyle=\small]
class MonteCarloTask implements Runnable { 
  private ArgumentGenIntf _randomVectorGenerator;
  ...
  public void run() {
    ...
    for (int i=0; i<_numTries; i++) {
      double[] xi=_randomVectorGenerator.
                   newVector(_lowerBnd, _upperBnd);
      double yi=_function.evaluate(xi);
      if (yi<local_incumbent_val) { 
        updateLocalIncumbent(xi,yi);
      }
    }
    // update global incumbent with local incumbent
    ...
  }
}

class DblArr1RndGenerator implements ArgumentGenIntf {
  public double[] newVector(double[] low, double[] high) {
    double[] x = new double[low.length];
    // populate x's elements
    ...
    return x;
  }
}
\end{lstlisting}

Using this approach, a commodity laptop equipped with an Intel core i7 CPU with 4 real cores with hyper-threading for a total of 8 logical processors and 16GB RAM, will run 10,000,000 evaluations of the 1000-dimensional Rastrigin multi-modal test-function~\cite{Muhlenbein} using 8 threads in approximately 202 seconds of wall-clock time. Does it matter that the method \texttt{newVector()} constructs a new \texttt{double[]} (and initializes its elements to zero) at every iteration? According to the “pools-are-bad” camp, the construction of these arrays should not be expected to hurt performance, and it results in a clean-looking code. Nevertheless, let’s add a private \texttt{double[] \_x;} simple member field that will act as a cache for a –now only once– constructed array, and modify the \texttt{newVector()} method as shown in Listing~\ref{lst:lst2}. As long as each thread has its own \texttt{DblArr1RndGeneratorWCache} object, there is no interference between the threads at all, and there is no need for sharing anything between the threads except for the final update of the global incumbent object when every thread will attempt to update the global incumbent with its own version of the best point found in its execution.

\begin{lstlisting}[caption={A slightly modified argument generator class that does not create new double arrays in each invocation of the \texttt{newVector()} method.},label={lst:lst2},captionpos=b,frame=single,basicstyle=\small]
class DblArr1RndGeneratorWCache implements ArgumentGenIntf {
  private double[] _x;
  ...
  public double[] newVector(double[] low, double[] high) {
    if (_x==null || low.length!=_x.length) 
      _x = new double[low.length];
    // populate _x elements
    ...
    return _x;
  }
}
\end{lstlisting}

\begin{figure}
  \centering
  \begin{minipage}[t]{0.45\textwidth}
    \includegraphics[width=\textwidth]{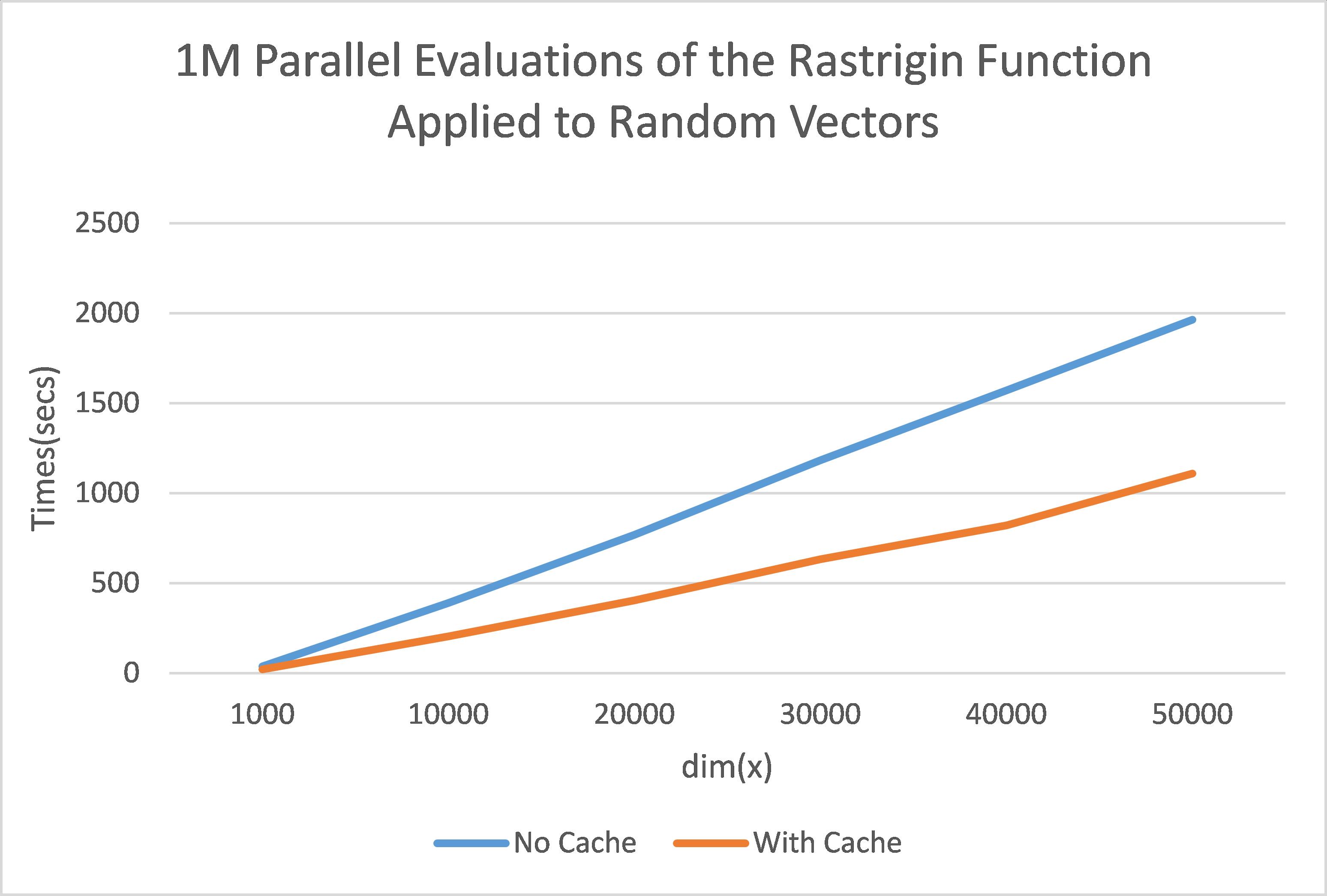}
    \caption{Comparing the performance of using or not a vector cache to evaluate the Rastrigin function in varying dimensions.}
    \label{lbl:fig1a}
  \end{minipage}
  \hfill
  \begin{minipage}[t]{0.45\textwidth}
    \includegraphics[width=\textwidth]{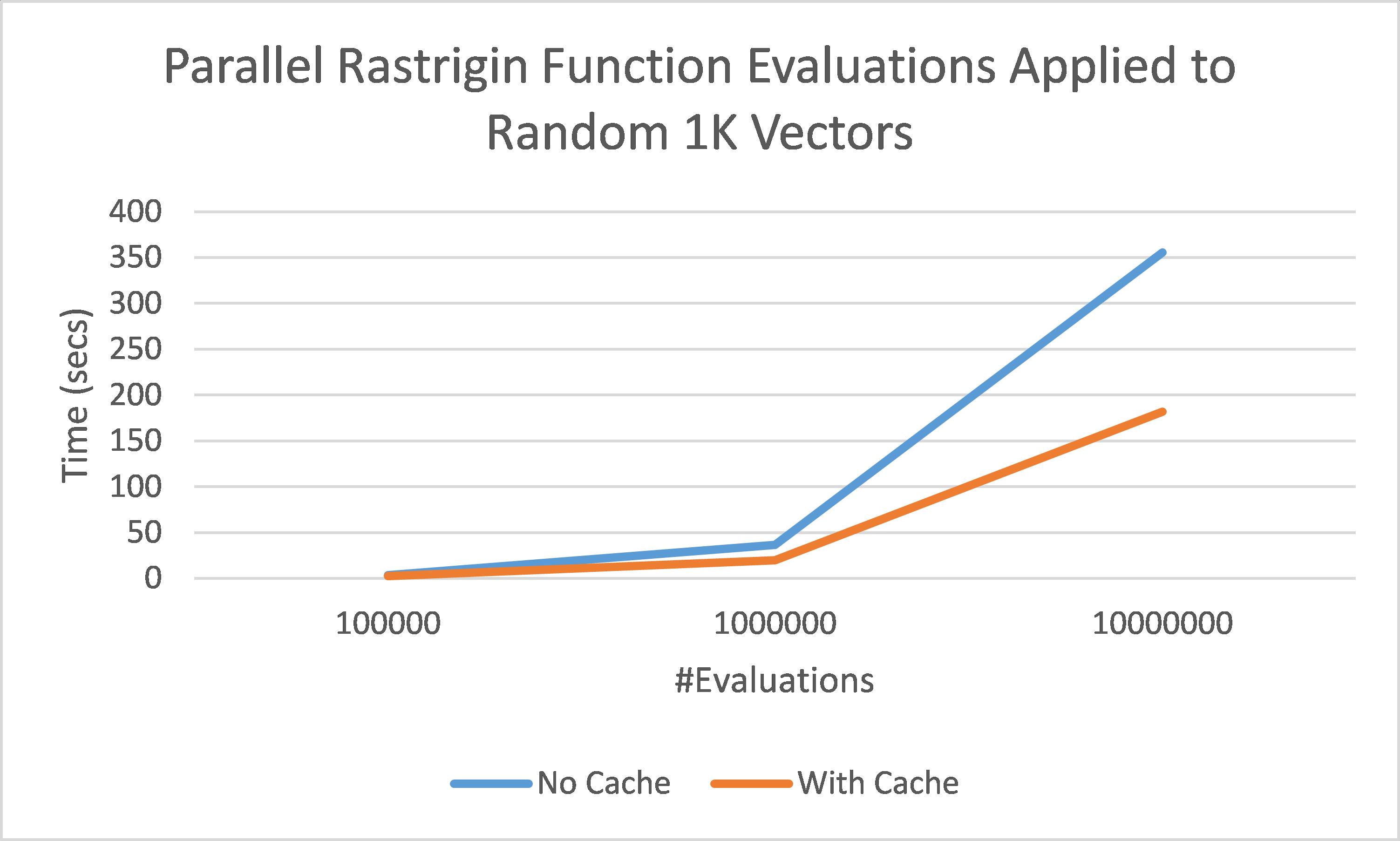}
    \caption{Comparing the performance of using or not a vector cache to evaluate the Rastrigin function in 1000 dimensions for various number of times. Both experiments conducted using 4 threads on an i7-3630QM CPU with 16GB RAM running Windows 10 Pro, compiled and run in JDK7.}
    \label{lbl:fig1b}
  \end{minipage}
\end{figure}

The code above now runs on the same hardware and the same number of threads in approximately 130 seconds. In fact, Fig.~\ref{lbl:fig1a} shows the running times when executing 1 million evaluations of the Rastrigin function in different number of dimensions, with and without this “thread-local member cache". One might argue that creating objects with sizes of up to 3.2 MB, doesn’t count as “light-weight object creation”. For this reason, we investigate the difference in running times when running a varying number of function evaluations for the 1000-dimensional Rastrigin function (which results in the creation of \texttt{double[]} objects of size 64K each, which by today's standards are usually considered quite “light-weight”) with and without the use of thread-local object pools. The results are shown in Fig.~\ref{lbl:fig1b} and again, show a rather strong advantage of this “data-member-as-pool” approach.

From the above it is clear that at least in some cases, taking care not to needlessly allocate new objects can prove beneficial for the overall system performance, though the reasons why this might be so may not be clear at this point. Next, we illustrate a similar case from a production system.

\section{Creating Many Small Objects In Recommender Systems}

Apache Mahout~\cite{Owen} defines the interface in Listing~\ref{lst:lst3} for recommended items.

\begin{center}
\begin{minipage}{\linewidth}
\begin{lstlisting}[caption={Recommended items interface in Apache Mahout.},label={lst:lst3},captionpos=b,frame=single,basicstyle=\small]
public interface RecommendedItem {
  long getItemID();
  float getValue();
}
\end{lstlisting}
\end{minipage}
\end{center}

This interface encapsulates a pair of values: an item-id together with its perceived value (that will be different for different users). Recommender engines, provide public interface methods as part of their public API that return a \texttt{List<RecommendedItem>} when asked to recommend a list of items to a specific user for example; when a recommender system engine follows this API style, it is clear that a huge number of very light-weight (and short-lived) memory objects holding such <id,value> pairs will be needed during a batch process of computing such recommendations for the many thousands of users subscribing to whatever service needs recommendations for, and exporting them to the engine clients. AMORE~\cite{Christou,Amolochitis} is a hybrid parallel recommender engine that the first author designed and implemented for a major Greek Triple-Play service provider, and its only means of interaction with the provider's systems is through a set of exposed web-services. AMORE runs periodically a batch process that computes for each of the more than forty thousand subscribers to the Video-Club service, the top-500 recommended content items. During this batch process, three different recommender algorithms (one user-based, one item-based, and one content-based) compute in parallel (using multiple concurrent threads each) the relative value of each of the available content items for each user, resulting in a total need of 3 $\times$ \#users $\times$ \#items $<$id,value$>$ pairs, which is in the order of $2\times 10^9$ objects created through multiple concurrently executing threads of control in a single iteration of the overall process. All these objects are very short-lived: as soon as the top-500 items for each subscriber have been computed, they are written to an RDBMS table, or sent through the network (XML encoded) to the consumer of an appropriate web-service method invoked). Such huge object generation requirements turn out to exercise serious stress on the JVM especially when running under strict memory requirements that force its garbage collector to run frequently, as is the case when a particular software system runs as a process on a virtual machine created by virtualization software hosted on server hardware that partitions the server’s physical resources among many several virtual machines, allocating a small fraction of the overall available RAM to each. 

The above discussion set the context in which our algorithms operate. To isolate from the rest of the system and study the effects of object creation and the possible benefits/trade-offs of using a thread-local object pool, we create a very simple benchmark program comprising of the two classes shown in Listings~\ref{lst:lst4} and~\ref{lst:lst5}.

\begin{lstlisting}[caption={No-pools implementation of a “holder” class holding <item,value> pairs.},label={lst:lst4},captionpos=b,frame=single,basicstyle=\small]
public class DLPair implements Serializable, RecommendedItem {
  private float _val; private long _id;
  public DLPair(float val, long id) { _val = val; _id = id; }
  public float getValue() { return _val; }
  public long getItemID() { return _id; }
  public void setData(long id, float v) { _id=id; _val=v; }
  public int compareTo(Object other) { ... }
}
\end{lstlisting}

\begin{lstlisting}[caption={Test-driver class testing the concurrent generation of many light-weight DLPair objects using the standard Java mechanism for new object creation (see line 46).},label={lst:lst5},captionpos=b,frame=single,basicstyle=\small]
public class DLPairTest {
  public static void main(String[] args) {
    long start = System.currentTimeMillis();
    long total_num_objs = Long.parseLong(args[0]);
    int num_threads = Integer.parseInt(args[1]);
    long start_ind = 0;
    long work_per_thread = total_num_objs/num_threads;
    DLPThread[] tds = new DLPThread[num_threads];
    for (int i=0; i<num_threads-1; i++) {
      tds[i] = new DLPThread(i, start_ind, work_per_thread);
      start_ind += work_per_thread;
    }
    long rem = total_num_objs - start_ind;
    tds[tds.length-1]=new DLPThread(tds.length-1,start_ind, rem);
    for (int i=0; i<tds.length; i++) {
      tds[i].start();
    }
    double total_val = 0.0;
    for (int i=0; i<tds.length; i++) {
      try {
        tds[i].join();
        total_val += tds[i].getSum();
      }
      catch (InterruptedException e) {
        Thread.currentThread().interrupt();
      }
    }
    long dur = System.currentTimeMillis() - start;
    System.out.println("Totalval="+total_val+" in "+dur+" msecs.");
  }
}
class DLPThread extends Thread {
  private int _id;
  private long _numObjs;
  private long _offset;
  private double _sum;
  public DLPThread(int id, long offset, long numobjs) {
    _numObjs = numobjs; _offset = offset; _id = id;
  } 
  public void run() {
    DLPair[] pis = new DLPair[10000];
    _sum = 0.0;
    int ind = 0;
    final long up_to = _offset+_numObjs;
    for (long i=_offset; i<up_to; i++) {
46:   DLPair pi = new DLPair(0, 0);
      pi.setData(i, 2.0f*i);  // do some computation
      _sum += pi.getValue();
      pis[ind++] = pi;
50:   if (ind==pis.length) ind=0;
    }
  }
  public double getSum() { return _sum; }
}
\end{lstlisting}

DLPairTest is a bare-bones implementation of a multi-threaded application that spawns a number of threads, each of which in its run method has a for-loop that iterates for a number of times, generating a DLPair object, setting values for the object’s members, and then, adding its value member to a running thread-local sum. The object is stored in an array, whose only reference is stack-allocated, until 10,000 more such objects are allocated in the for-loop, at which time it is removed from this array and becomes available for garbage-collection. The main program waits until each of its spawned threads are joined and then exits. Notice that in Listing~\ref{lst:lst5}, creation of each object occurs via calling \texttt{new DLPair(.,.)} in line 46. The program, when executed on the same machine mentioned as above, with the JVM maximum heap option \texttt{-Xmx100m} specified, and parameters 800,000,000 8 (i.e., 800 mio objects to be created in equal numbers by 8 threads) runs in approximately 5.8 seconds of wall-clock time, as averaged over 10 runs. As it turns out, we can improve substantially the performance of this program by using the concept of a completely lock-free thread-local object pool that essentially acts as a convenient cache of frequently allocated objects in thread-local storage. The important requirement is that the program flow must be such so that objects “requested” in one thread, essentially remain visible only in the same thread that requested them, and are eventually “released” back into the pool from the same thread that requested them in the first place.

\section{A Generic Thread-Local Object Pool API}

Consider an API for poolable objects that can be stored in a generic thread-local object cache (which we appropriately call, \texttt{ThreadLocalObjectPool}) as shown in Listing~\ref{lst:lst6}. Once a software designer decides that a particular class X would benefit from thread-local object pooling, they have to declare that X extends the \texttt{PoolableObject} class, and implement the method \texttt{setData(Object... args)}. They also need to modify the constructors of class X to call \texttt{super(null)} first, and add a constructor taking a pool as an argument, with a body of calling \texttt{super(pool)}.

\begin{lstlisting}[caption={The abstract \texttt{PoolableObject} base-class.},label={lst:lst6},captionpos=b,frame=single,basicstyle=\small]
public abstract class PoolableObject implements Serializable {
  private ThreadLocalObjectPool _pool = null;
  private boolean _isUsed = false;
    
  public final static <T extends PoolableObject> T 
    newInstance(PoolableObjectFactory<T> f, Object... args) {
    T t = ThreadLocalObjectPool.<T>getObject(f, args);
    t.setData(args);
    return t;
  }
    
  protected PoolableObject(ThreadLocalObjectPool pool)
  { _pool = pool; }
    
  /**
   * called by <CODE>newInstance(f,args)</CODE> method when 
   * returning a new instance of the implementing sub-class.
   * @param args 
   */
  public abstract void setData(Object... args);
    
  final public void release() {
    if (_pool!=null) {
      if (_isUsed) {
        _isUsed=false;
        _pool.returnObjectToPool(this);
      } else throw
        new IllegalStateException("Object not currently used");
    }
  }
  final protected void setIsUsed() { _isUsed = true; }
  final protected boolean isUsed() { return _isUsed; }
  final protected boolean isManaged() { return _pool!=null; }
}
\end{lstlisting}

Such an implementation is usually trivial: for the example of the \texttt{DLPair} class above, one would write the \texttt{setData} method like this:
\begin{verbatim}
public void setData(Object... args) {
  if (args==null) return;  // guard against null var-args
    _id = ((Long) args[0]).longValue();
    _val = ((Double) args[1]).doubleValue();
  }
\end{verbatim}

To use the newly defined \texttt{DLPair} objects, we also need an appropriate factory class implementing the following interface shown in Listing~\ref{lst:lst7}.

\begin{lstlisting}[caption={The \texttt{PoolableObjectFactory} public interface.},label={lst:lst7},captionpos=b,frame=single,basicstyle=\small]
/**
 * interface for factory creating objects of a class whose
 * instances may be  "managed" or "unmanaged" by thread-local 
 * object pools; such classes must extend the
 * <CODE>PoolableObject</CODE> base abstract class.
 */
public interface PoolableObjectFactory<T extends PoolableObject> {
  /**
   * compile-time constant specifies the maximum number of types
   * that will need thread-local pooling in the program.
   */
  public static final int _MAX_NUM_POOLED_TYPES = 10;

  /**
   * create an "unmanaged" object, using the args passed in.
   * @param args
   * @return 
   */
  T createObject(Object... args);

  /**
   * create a "managed" object (using args passed in) that will
   * belong to the specified pool.
   * @param pool
   * @param args
   * @return 
   */
  T createPooledObject
    (ThreadLocalObjectPool<T> pool, Object... args);

  /**
   * return an id that is unique for each "poolable" type T
   * in the program.
   * This id will be used to dereference the right pool 
   * of <CODE>PoolableObject</CODE>s existing in the
   * <CODE>ThreadLocal</CODE> variable of the current thread
   * when getting an object of type T. 
   * @return int must be within {0,..._MAX_NUM_POOLED_TYPES}.
   */
  int getUniqueTypeId();
}
\end{lstlisting}

The \texttt{ThreadLocalObjectPool} generic class is illustrated in Listing~\ref{lst:lst8}. The factory method \texttt{getUniqueTypeId()} deserves some special mention as it might not be immediately clear what purpose it serves. Given that it is quite possible that in a program there may be a need for more than one type of object to be pooled, in order to quickly fetch from thread-local storage the pool for a particular type, for performance reasons, we use a thread-local array of pools (see Listing~\ref{lst:lst9}); the targeted array’s position is identified by the value the \texttt{getUniqueTypeId()} method returns. The alternative of identifying the correct 
pool to fetch by the name of the class of the objects we need incurs high performance penalties. Here is why: in such a design, an associative array (a map) of a kind would be needed to store the thread-local object pool of each object type by name; but the \texttt{get()} operation of maps that would then be required in each invocation of the \texttt{newInstance()} method of any \texttt{PoolableObject} which is much slower than the simple array de-reference operation that is required in the current design.

Returning to the example DLPairTest program, an example factory creating pooled DLPair objects becomes as simple as this:

\begin{verbatim}
class DLPairFactory implements PoolableObjectFactory<DLPair> {
  public DLPair createObject(Object... args) {
    return new DLPair(null);
  }
  public DLPair createPooledObject
         (ThreadLocalObjectPool<DLPair> pool, Object... args) {
    return new DLPair(pool);
  }
  public int getUniqueTypeId() { return 0; }
}
\end{verbatim}

\begin{lstlisting}[caption={The \texttt{ThreadLocalObjectPool} generic class that can be used to hold any \texttt{PoolableObject}.},label={lst:lst8},captionpos=b,frame=single,basicstyle=\small]
public final class ThreadLocalObjectPool<T extends PoolableObject> 
  implements Serializable {
  // default cache size
  private static volatile int _NUMOBJS = 100000;
    
  private T[] _pool;  // the container of the objects
  private int _topAvailPos = -1;
    
  static <T extends PoolableObject> ThreadLocalObjectPool<T> 
    newThreadLocalObjectPool(PoolableObjectFactory<T> f,
                             Object... args) {
    ThreadLocalObjectPool<T> p = new ThreadLocalObjectPool<>();
    p.initialize(f, args);
    return p;
  }
    
  private ThreadLocalObjectPool() {
    _pool = (T[]) new PoolableObject[_NUMOBJS];
  }
    
  static <T extends PoolableObject> 
    T getObject(PoolableObjectFactory<T> f, Object... args) {
    ThreadLocalObjectPool<T> pool = 
      ThreadLocalObjectPools.<T>getThreadLocalPool(f, args);
    T p = pool.getObjectFromPool();
    if (p!=null) {  // ok, return managed object
      return p;
    } else  // oops, create new unmanaged object
      return f.createObject(args);
  }
    
  void returnObjectToPool(T ind) { _pool[++_topAvailPos] = ind; }
  
  private void 
    initialize(PoolableObjectFactory<T> f, Object... args) {
    for (int i=0; i<_NUMOBJS; i++) {
      _pool[i] = f.createPooledObject(this,args);
    }
    _topAvailPos = _NUMOBJS-1;        
  }

  private T getObjectFromPool() {
    if (_topAvailPos>=0) {
      T obj = _pool[_topAvailPos--];
      _pool[_topAvailPos+1] = null;  // avoid memory leaks
      obj.setIsUsed();
      return obj;
    }
    return null;
  }
	
  static void setPoolSize(int num) {...} 
  static int getPoolSize() { return _NUMOBJS; }
}
\end{lstlisting}

\begin{center}
\begin{minipage}{\linewidth}
\begin{lstlisting}[caption={The \texttt{ThreadLocalObjectPools} auxiliary class.},label={lst:lst9},captionpos=b,frame=single,basicstyle=\small]
public final class ThreadLocalObjectPools {
  private static boolean _poolSizeResetAllowed = true;  
  private static ThreadLocal<ThreadLocalObjectPool[]> _poolsMap = 
    new ThreadLocal();
  
  public static <T extends PoolableObject> void 
    deleteThreadLocalPool(PoolableObjectFactory<T> f) { ... }
        
  static <T extends PoolableObject> ThreadLocalObjectPool<T>
    getThreadLocalPool(PoolableObjectFactory<T> f, Object... ars) 
    {
      ThreadLocalObjectPool[] p = 
        (ThreadLocalObjectPool[]) _poolsMap.get();
      ThreadLocalObjectPool<T> pool = null;
      if (p==null) {
        synchronized (ThreadLocalObjectPools.class) {
          _poolSizeResetAllowed=false;
        }
        p = new ThreadLocalObjectPool[PoolableObjectFactory.
                  _MAX_NUM_POOLED_TYPES];
        _poolsMap.set(p);
      }
      final int fid = f.getUniqueTypeId();
      pool = p[fid];
      if (pool==null) {
        pool = 
          ThreadLocalObjectPool.<T>newThreadLocalObjectPool(f,ars);
        p[fid] = pool;
      }   
      return pool;              
    }
            
  static <T extends PoolableObject> ThreadLocalObjectPool<T>
    getThreadLocalPool(PoolableObjectFactory<T> f) 
      throws IllegalStateException { ... } 

  public static synchronized void setPoolSize(int poolsize) 
    throws IllegalStateException, IllegalArgumentException { ... }

  public static int getPoolSize() { 
    return ThreadLocalObjectPool.getPoolSize();
  }
}
\end{lstlisting}
\end{minipage}
\end{center}

\begin{figure}
  \centering
  \begin{minipage}[t]{0.45\textwidth}
    \includegraphics[width=\textwidth]{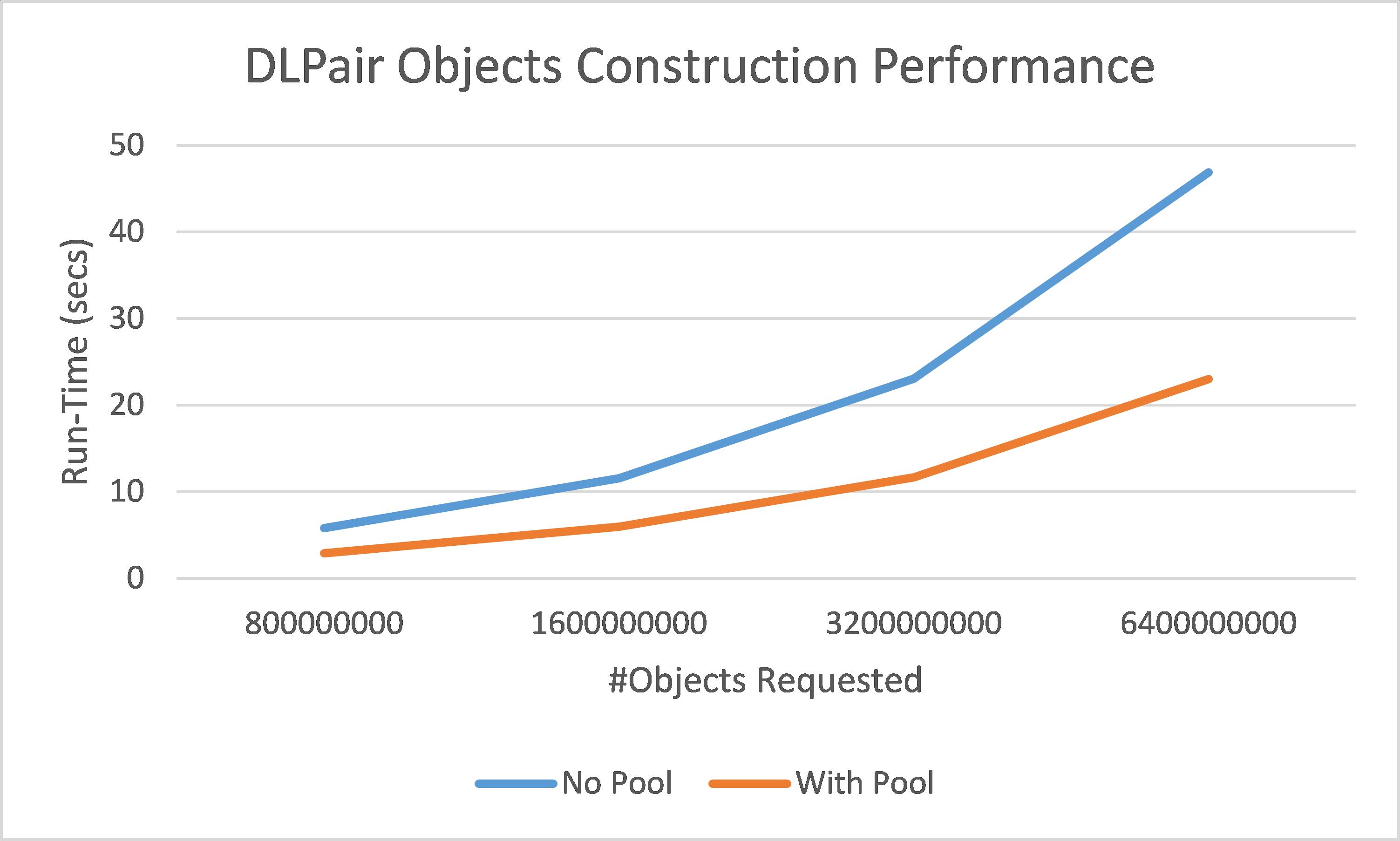}
    \caption{Performance comparison of a sample test-driver program constructing large numbers of trivially small objects, concurrently on 8 threads, under JVM memory constraint “-Xmx100m”. }
    \label{lbl:fig2a}
  \end{minipage}
  \hfill
  \begin{minipage}[t]{0.45\textwidth}
    \includegraphics[width=\textwidth]{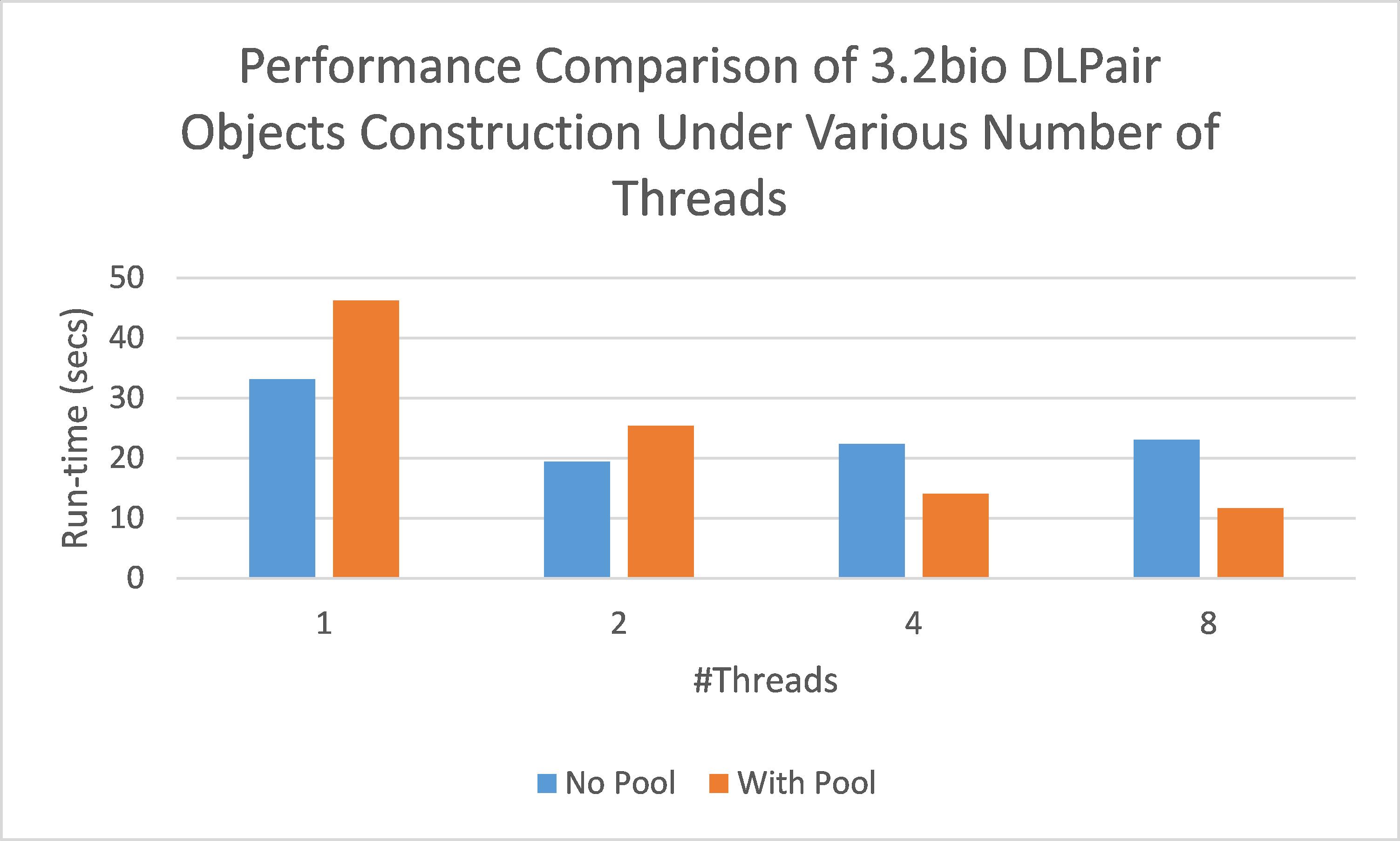}
    \caption{Comparing the performance of using Thread-Local Object Pool under various degrees of concurrency and under same JVM memory constraint as in Figure~\ref{lbl:fig2a}.}
    \label{lbl:fig2b}
  \end{minipage}
\end{figure}

Finally, we modify line 50 to read as follows:

\begin{verbatim}
50: if (ind == pis.length) {
        for (int j=0; j<pis.length; j++) pis[j].release();
        ind = 0;
    }
\end{verbatim}

With this set of changes, the \texttt{DLPairTest} sample code now runs in 2.9 seconds. This performance gain is consistent as the number of objects and/or number of threads grows, as seen in Figures~\ref{lbl:fig2a} and~\ref{lbl:fig2b}.

\begin{figure}
  \centering
  \begin{minipage}[t]{0.45\textwidth}
    \includegraphics[width=\textwidth]{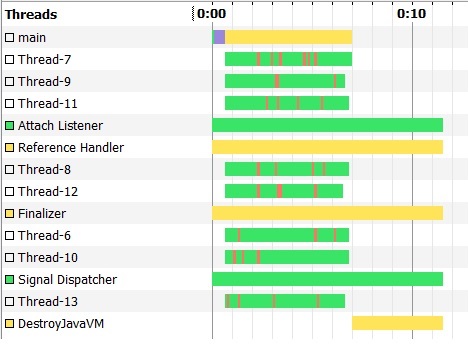}
    \caption{Profile of a sample test-driver program constructing large numbers of trivially small objects, concurrently on 8 threads, without the use of pooling, under JVM memory constraint “-Xmx100m”. All user threads experience blocking even though there is no explicit user-lock in the program.}
    \label{lbl:fig3a}
  \end{minipage}
  \hfill
  \begin{minipage}[t]{0.45\textwidth}
    \includegraphics[width=\textwidth]{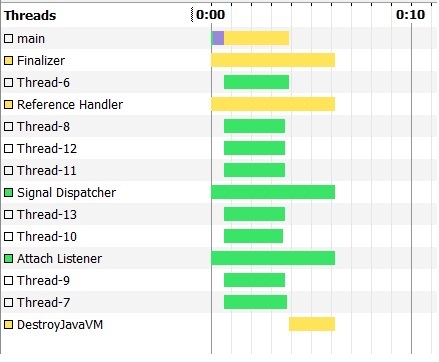}
    \caption{Profiling the performance of the same test-driver program as in Figure~\ref{lbl:fig3a} but with 
    the use of pooling. The threads never stall.}
    \label{lbl:fig3b}
  \end{minipage}
\end{figure}

Where does the gain come from? In Figures~\ref{lbl:fig3a} and~\ref{lbl:fig3b} we show a profiling of running the 
same program, with and without pooling. It is clear that when pooling is not used, all the participating threads
experience some locking, even though the program itself is lock-free. This locking is necessitated by the need of
the threads to allocate more memory and/or by the garbage collector that kicks in while the program is running.

\begin{figure}
\centering
\includegraphics[width=.4\textwidth]{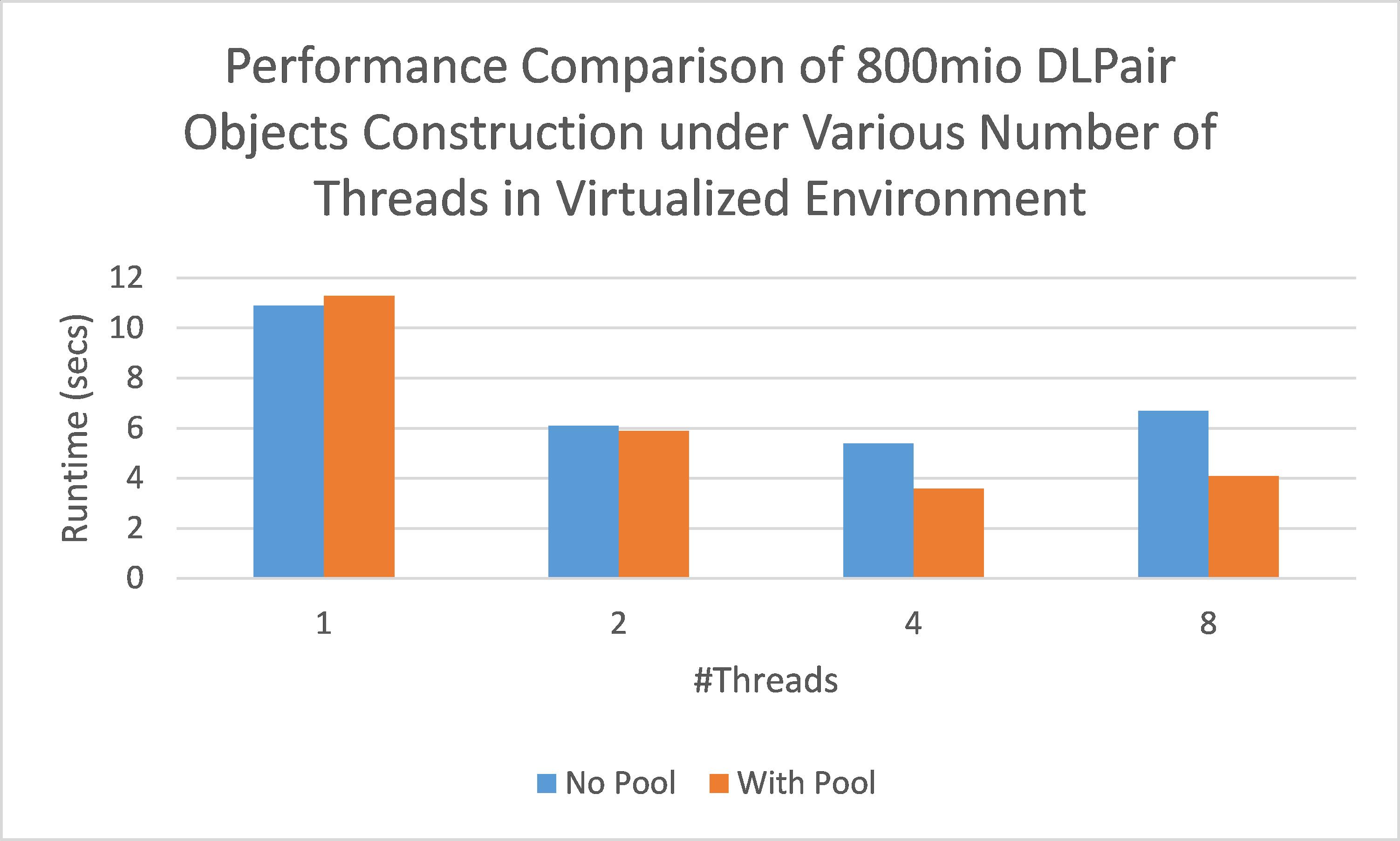}
\caption{Thread-local object pool performance in a virtualized environment running OpenMandriva Linux 2014 64-bit with access to 4 CPU cores, and Java8 running under \texttt{-Xmx100m} JVM option.}
\label{lbl:fig3}
\end{figure}

The gain observed when constructing many lightweight short-lived objects from many concurrently executing threads still exists when running in a virtualized environment: we executed the same program on an Oracle VirtualBox virtual machine running OpenMandriva Linux 2014 64-bit, hosted on the same hardware as our previous experiments, allowing the virtual machine to utilize 4 cores of the real CPU. The JVM running on the virtual machine was now JDK 8. The results shown in Figure~\ref{lbl:fig3} remain consistent with what we previously observed. Further, we also verified the trend of the thread-local object pool to out-perform the no-pool approach in the presence of multiple threads on different CPUs and Operating Systems; in particular, on an i3 dual-core CPU running Windows 8.1, and on an Apple MacBook Pro mid2014 with an i7 quad core CPU running OS X. We also compared the performance of our proposed Thread-Local Object Pool with that of the so-called “FastObjectPool”~\cite{FastObjectPool}, which is a lock-less object pool that has been measured to compare favorably against some other object pools available on the internet. Unfortunately, it appears to suffer from bugs that often cause it to loop endlessly, or at least to be more than 4 orders of magnitude slower, in the presence of large numbers of object creation requests.

\section{Discussion and Conclusions}

We show that thread-local object pools acting as caches for objects that are usually not shared among different threads can significantly improve the performance of computationally intensive tasks, especially under heavy concurrency and stringent memory requirements, even though their performance can sometimes lag behind the no-pool-approach in single-threaded environments. Creating many light-weight (and usually short-lived) objects, is often an indication of sub-optimal data design, or application of Object-Oriented concepts at too fine a level of granularity. In such cases, whenever the software designer has control over such interface choices, it would make much more sense to group data in more meaningful ways for performance, rather than try to alleviate the issues caused by fine granularity of objects by using object pools on top. But whenever such design choices cannot be altered, and one has to deal with the fact that large numbers of lightweight objects will have to be requested on demand from many concurrently executing threads, then using a small library as the one we propose can result in significant performance gains. The proposed code for the library we describe can be found as Open-Source at \texttt{https://github.com/ioannischristou/ThreadLocalObjectPool}.

\end{document}